\documentclass[a4paper]{jpconf}
\usepackage{citesort}
\usepackage{graphics}
\usepackage{graphicx}
\newcommand{\slim}{\mskip 1.5mu}              
\newcommand{\lf}{\left}
\newcommand{\rg}{\right}
\begin{document}
\title{Transverse spin dependent azimuthal asymmetries at COMPASS}
\author{B.~Parsamyan$^{1,}$ $^2$}
\address{$^1$ Dipartimento di Fisica Generale, Universit\`a di Torino, Torino, Italy}
\address{$^2$ INFN, Sezione di Torino, Via P. Giuria 1, I-10125 Torino, Italy}
\ead{bakur.parsamyan@cern.ch}
\begin{abstract}
In the semi-inclusive deep inelastic scattering of polarized leptons
on a transversely polarized target eight target transverse
spin-dependent azimuthal modulations are allowed. In the QCD parton
model half of these asymmetries can be interpreted within the
leading order approach and the other four are twist-three
contributions.  The first two leading twist asymmetries extracted by
HERMES and COMPASS experiments are related: one to the transversity
distribution and the Collins effect, the other to the Sivers
distribution function. These results triggered a lot of interest in
the past few years and allowed the first extractions of the
transversity and the Sivers distribution functions of nucleon. The
remaining six asymmetries were obtained by the COMPASS experiment
using a 160 GeV/c longitudinally polarized muon beam and
transversely polarized deuteron and proton targets. Here we review
preliminary results from COMPASS proton data of 2007.
\end{abstract}
\section{Introduction}\label{intro}

Using standard SIDIS notations transverse spin $S_T$ dependent part
of the general, model-independent cross-section of lepton-hadron
SIDIS processes can be written in the following way
\cite{Kotzinian:1994dv}, \cite{Bacchetta:2006tn}:
\begin{eqnarray}\label{e:crossmaster}
&&\frac{d\sigma}{dx \, dy\, d\psi \,dz\, d\phi_h\, d
P_{hT}^2} = \frac{\alpha^2}{x y\slim
Q^2} \frac{y^2}{2(1-\varepsilon)} \biggl( 1+\frac{\gamma^2}{2x}
\biggr)\Biggl\{... \\
&&  +|{\bf S}_T| \Bigg[ \sin(\phi_h-\phi_S) \Bigl(F_{UT
,T}^{\sin\lf(\phi_h -\phi_S\rg)} + \varepsilon
F_{UT,L}^{\sin\lf(\phi_h -\phi_S\rg)}\Bigr)+ \varepsilon \sin(\phi_h+\phi_S) F_{UT}^{\sin\lf(\phi_h +\phi_S\rg)}\nonumber\\
&& + \varepsilon \sin(3\phi_h-\phi_S) F_{UT}^{\sin\lf(3\phi_h -\phi_S\rg)}
 + \sqrt{2\varepsilon (1+\varepsilon)}\biggl( \sin\phi_S F_{UT}^{\sin \phi_S } +
\sin(2\phi_h-\phi_S) F_{UT}^{\sin\lf(2\phi_h -\phi_S\rg)}\biggl) \Bigg] \nonumber\\
&& + |{\bf S}_T| P_{l} \Bigg[\sqrt{1-\varepsilon^2} \cos(\phi_h-\phi_S) F_{LT}^{\cos(\phi_h -\phi_S)}
+ \sqrt{2\varepsilon (1-\varepsilon)} \cos\phi_S F_{LT}^{\cos \phi_S} \nonumber \\
&& + \sqrt{2\varepsilon (1-\varepsilon)}\cos(2\phi_h-\phi_S)
F_{LT}^{\cos(2\phi_h - \phi_S)}\Bigg] \Biggr\}\nonumber,\quad \mathrm{where} \,\,
\varepsilon = \frac{1-y -\frac{1}{4}\slim \gamma^2 y^2}{1-y +\frac{1}{2}\slim y^2 +\frac{1}{4}\slim \gamma^2 y^2}, \,\,\gamma = \frac{2 M x}{Q}
\end{eqnarray}
This expression contains eight azimuthal modulations in the $\phi_h$
and $\phi_S$ (azimuthal angles of the produced hadron and of the
nucleon spin correspondingly). Each modulation leads to an asymmetry
described by the associated structure function $F$ depending on $x$,
$Q^2$, $z$ and $P_{hT}$. The superscript of the structure function
indicates the corresponding modulation, the first and the second
subscripts - the respective ("U"-unpolarized,"L"-longitudinal and
"T"-transverse) polarization of beam and target and the third
subscript specifies the polarization of the virtual photon. The
eight Transverse Spin Asymmetries (TSA) linked to the azimuthal
modulations are defined as the ratios of the corresponding structure
functions to the unpolarized one $A_{BT}^{w_i(\phi_h, \phi_s)}
\equiv F_{BT}^{w_i(\phi_h,\phi_s)}/F_{UU,T}$ where $B=L$ or $B=U$
indicates the beam polarization. As one can see, there are five
Single-Spin Asymmetries (SSA), which depend only on $S_T$ and three
Double-Spin Asymmetries (DSA), both $S_T$ and $P_{l}$ (beam
polarization) dependent. In the QCD parton model approach four of
the eight TSAs have a Leading Order (LO) interpretation and are
described by the convolutions of Transverse Momentum Dependent (TMD)
twist-two distribution functions (DFs) and fragmentation
functions(FFs):
\begin{eqnarray}
&&A_{UT}^{\sin (\phi _h -\phi _s )} \propto f_{1T}^{\bot q} \otimes
D_{1q}^h,\ \
A_{UT}^{\sin (\phi _h +\phi _s )} \propto h_1^q \otimes H_{1q}^{\bot
h}, \\
&&A_{LT}^{\cos (\phi _h -\phi _s )} \propto g_{1T}^q \otimes
D_{1q}^h,\ \
A_{UT}^{\sin (3\phi _h
-\phi _s )} \propto h_{1T}^{\bot q} \otimes H_{1q}^{\bot
h}\nonumber
\label{eq:LO_as}
\end{eqnarray}
Here the first two are Collins and Sivers asymmetries, which have
been extracted from HERMES experimental data collected on proton
target \cite{Airapetian:2004tw,Airapetian:2009ti} and COMPASS
deuteron \cite{Alexakhin:2005iw,Ageev:2006da,Alekseev:2008dn} and
proton \cite{Alekseev:2010rw} data. These measurements triggered a
strong phenomenological and theoretical interest and, for example,
allowed the first extraction of the Sivers DF and combined with the
BELLE data \cite{Abe:2005zx,Seidl:2008xc} - of the transversity
function and Collins FF
\cite{Anselmino:2005ea,Anselmino:2007fs,Anselmino:2008jk}.

The other two $A_{LT}^{\cos (\phi _h -\phi _s )}$ and $A_{UT}^{\sin
(3\phi _h -\phi _s )}$ LO TSAs can be used for extraction of
$g_{1T}^q$ (worm-gear) and $h_{1T}^{\perp\,q}$ (pretzelosity) DFs
correspondingly. There are different model-based calculations of
these asymmetries made for COMPASS kinematics. In this article we
refer, for instance, to the following ones: $A_{LT}^{\cos (\phi _h
-\phi _s )}$ DSA calculated using "Kotzinian-Mulders" model
\cite{Kotzinian:1995cz,Kotzinian:2006dw}, and predictions for both
$A_{LT}^{\cos (\phi _h -\phi _s )}$ and $A_{UT}^{\sin (3\phi _h
-\phi _s )}$ based on light-cone constituent quark models
\cite{Boffi:2009sh,Pasquini:2008ax} and quark-diquark model
\cite{Kotzinian:2008fe}.

The remaining four asymmetries are higher-twist effects, though they
can be interpreted as twist-two Cahn kinematic corrections to spin
effects on the transversely polarized nucleon
\cite{Kotzinian:1994dv}:
\begin{eqnarray}
&&A_{UT}^{\sin (\phi _s )} \propto \frac{M}{Q}({h_1^q \otimes
H_{1q}^{\bot h} +f_{1T}^{\bot q} \otimes D_{1q}^h }),\ \
A_{UT}^{\sin (2\phi _h -\phi _s )} \propto
\frac{M}{Q}({h_{1T}^{\bot q} \otimes H_{1q}^{\bot h}
+f_{1T}^{\bot q} \otimes D_{1q}^h }) \nonumber\\
&&A_{LT}^{\cos (\phi _s )} \propto \frac{M}{Q}(g_{1T}^q \otimes
D_{1q}^h),\ \ A_{LT}^{\cos (2\phi _h -\phi _s )} \propto \frac{M}{Q}
(g_{1T}^q \otimes D_{1q}^h).
 \label{eq:subLO_as}
\end{eqnarray}
For instance, such an approach was used in \cite{Kotzinian:2008fe}
in order to evaluate these TSAs based on the quark-diquark model.

Both LO $A_{LT}^{\cos (\phi _h -\phi _s )}$ and $A_{UT}^{\sin (3\phi
_h -\phi _s )}$ as well as the other four $A_{UT}^{\sin (\phi _s
)}$, $A_{UT}^{\sin (2\phi _h -\phi _s )}$, $A_{LT}^{\cos (\phi _s
)}$ and $A_{LT}^{\cos (2\phi _h -\phi _s )}$ "higher-twist"
asymmetries have been extracted from COMPASS deuteron data
\cite{Parsamyan:2007ju,Kotzinian:2007uv} and, together with the
measurement of Collins and Sivers effects they complete the whole
"deuteron"-set of eight allowed in SIDIS TSAs. Now in the next
section we present our preliminary results for all six "beyond
Collins and Sivers" TSAs from COMPASS proton data.
\begin{figure}[ht!]
\center
\includegraphics[width=1\textwidth]{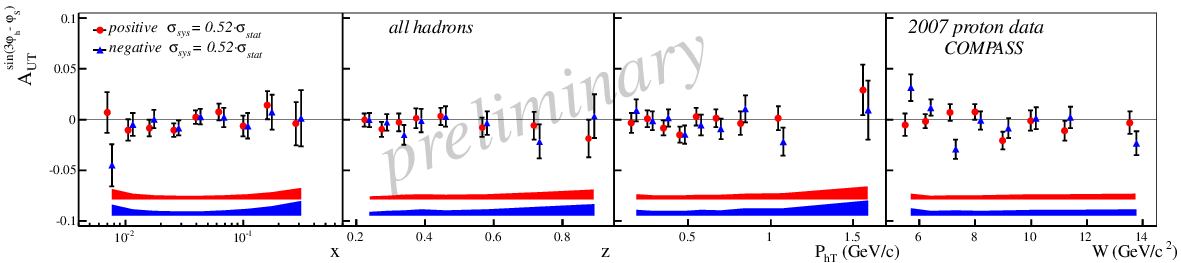}
\includegraphics[width=1\textwidth]{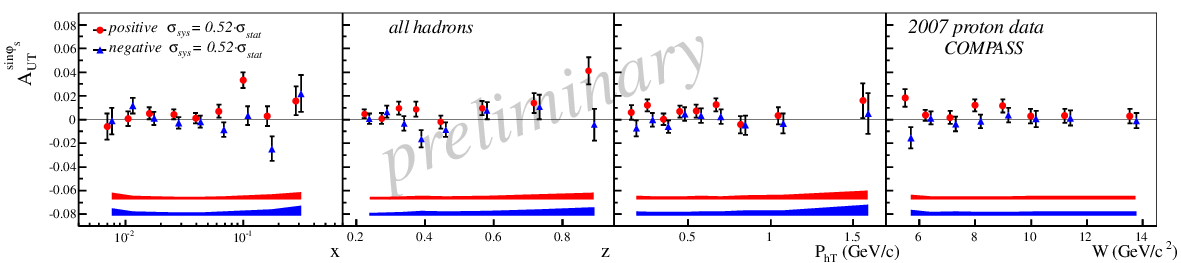}
\includegraphics[width=1\textwidth]{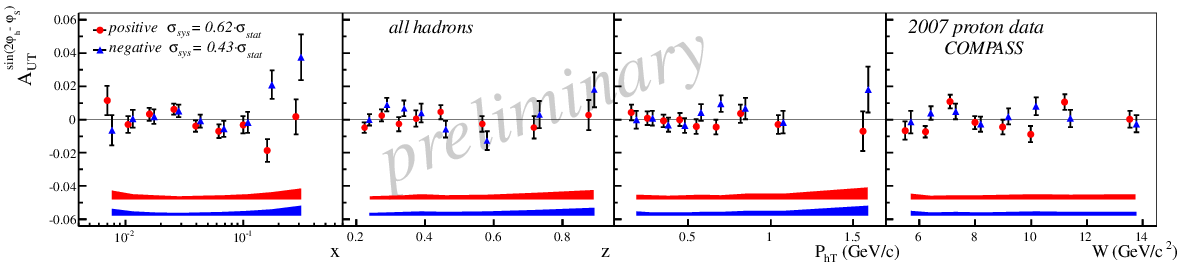}
\caption{$A_{UT}^{sin(3\phi_h-\phi_s)}$, $A_{UT}^{sin\phi_s}$ and $A_{UT}^{sin(2\phi_h-\phi_s)}$
asymmetries for positive (red circles) and negative (blue triangles)
hadrons vs. $\it x$, $z$, $P_{hT}$ and $W$. Systematic uncertainties are shown by the correspondingly colored bands.}
\label{fig:1}       
\end{figure}
\begin{figure}[ht!]
\center
\includegraphics[width=1\textwidth]{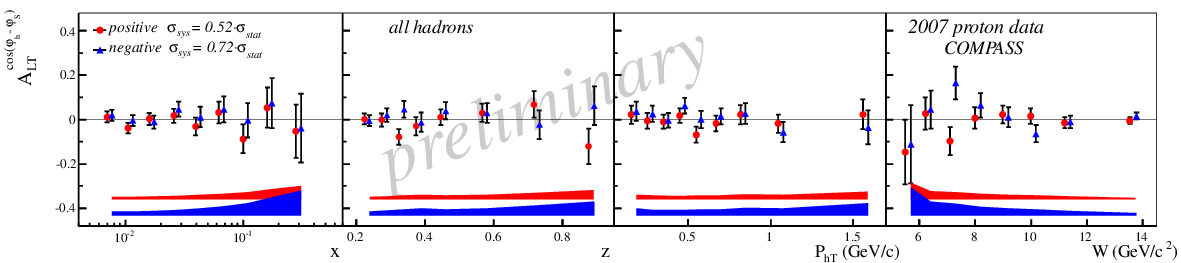}
\includegraphics[width=1\textwidth]{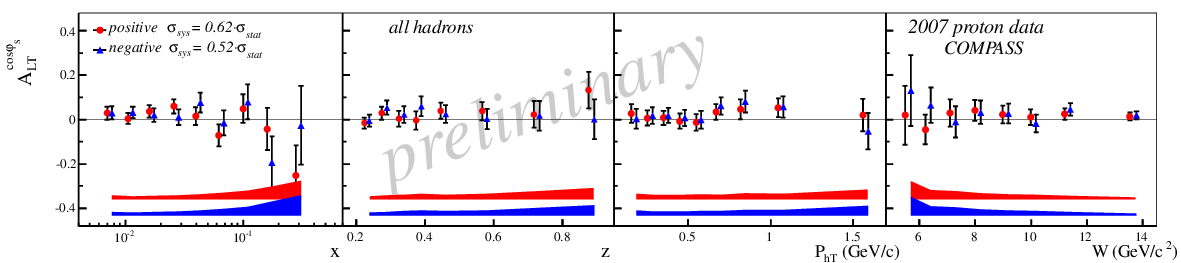}
\includegraphics[width=1\textwidth]{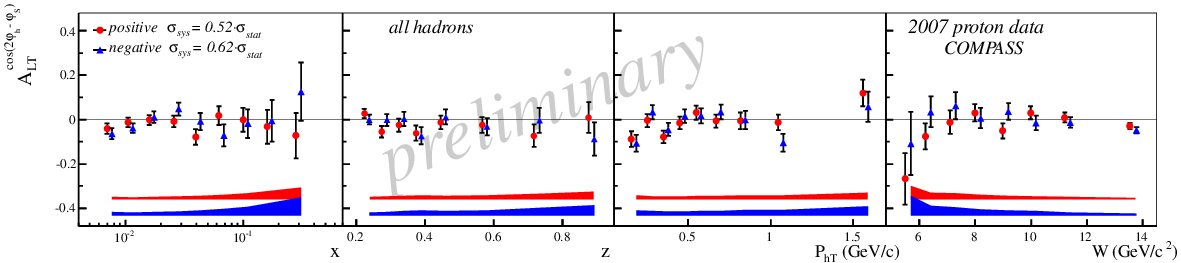}
\caption{$A_{LT}^{cos(\phi_h-\phi_s)}$, $A_{LT}^{cos\phi_s}$ and $A_{LT}^{cos(2\phi_h-\phi_s)}$
asymmetries for positive (red circles) and negative (blue triangles)
hadrons vs. $\it x$, $z$, $P_{hT}$ and $W$. Systematic uncertainties are shown by the correspondingly colored bands.}
\label{fig:2}       
\end{figure}
\begin{figure}[h!]
\center
\includegraphics[width=1\textwidth]{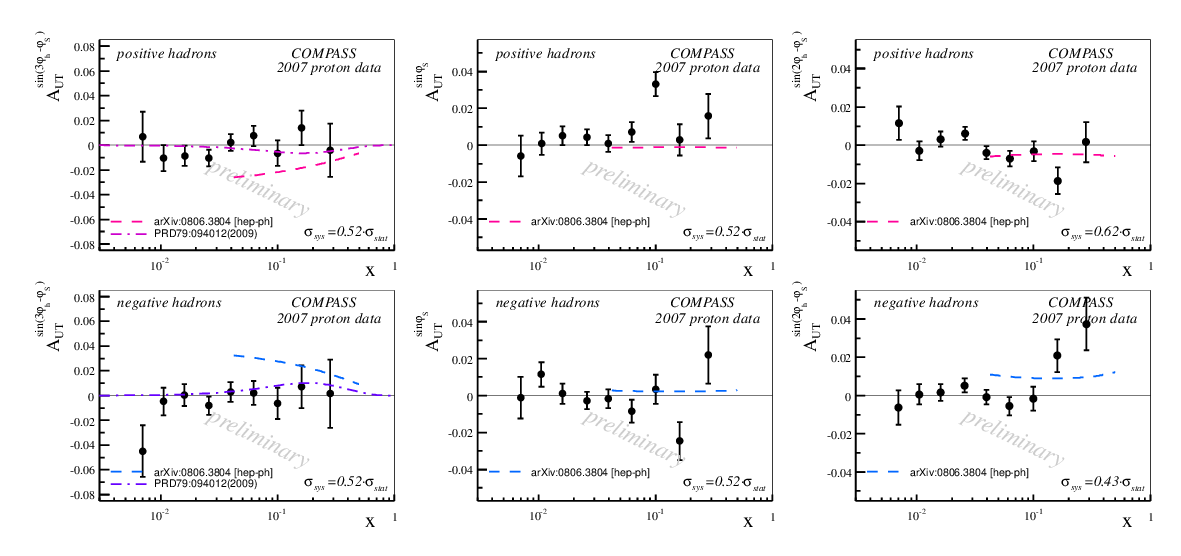}
\caption{$A_{UT}^{sin(3\phi_h-\phi_s)}$, $A_{UT}^{sin\phi_s}$, $A_{UT}^{sin(2\phi_h-\phi_s)}$
asymmetries for $h^{+}$ (top) and  $h^{-}$ (bottom) vs. $\it x$.
compared with the predictions from: \cite{Pasquini:2008ax} (dot-dashed) and \cite{Kotzinian:2008fe} (dashed).}
\label{fig:3}
\end{figure}
\begin{figure}[h!]
\center
\includegraphics[width=1\textwidth]{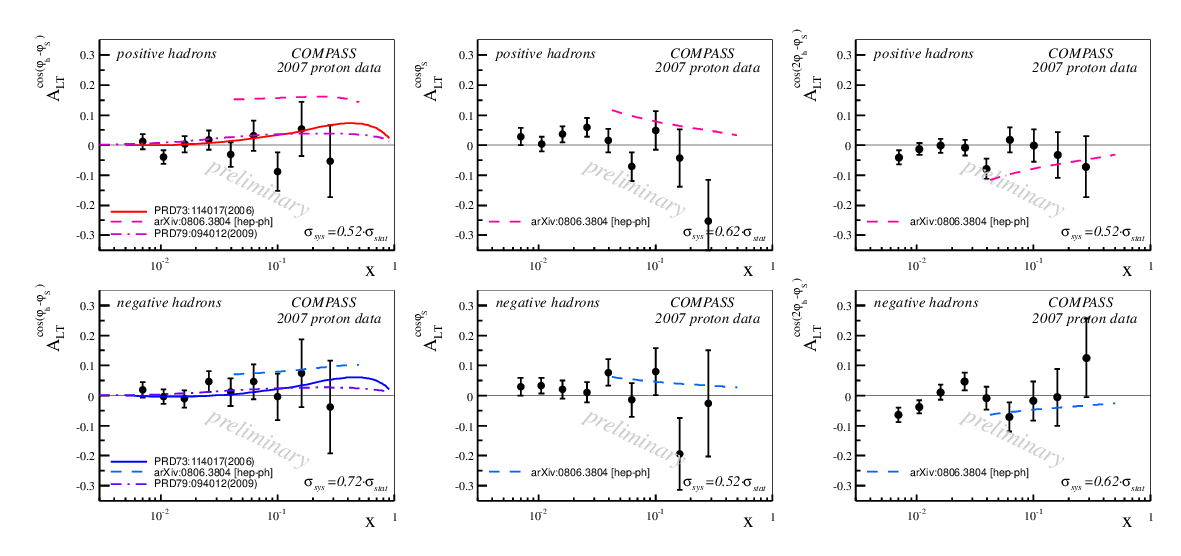}
\caption{$A_{LT}^{cos(\phi_h-\phi_s)}$,$A_{LT}^{cos\phi_s}$,$A_{LT}^{cos(2\phi_h-\phi_s)}$,
asymmetries for $h^{+}$ (top) and  $h^{-}$ (bottom) vs. $\it x$.
compared with the predictions from: \cite{Kotzinian:2006dw} (solid), \cite{Pasquini:2008ax} (dot-dashed) and \cite{Kotzinian:2008fe} (dashed).}
\label{fig:4}
\end{figure}
\section{Results}\label{Results}

In COMPASS transverse $A_{UT}^{\sin (3\phi _h -\phi _s )}$,
$A_{UT}^{\sin (\phi _s )}$ and $A_{UT}^{\sin (2\phi _h -\phi _s )}$
SSAs and $A_{LT}^{\cos (\phi _h -\phi _s )}$, $A_{LT}^{\cos (\phi _s
)}$ and $A_{LT}^{\cos (2\phi _h -\phi _s )}$ DSAs were extracted for
positive and negative hadron production in SIDIS of high energy
muons on transversely polarized protons. The data were collected in
2007 using 160 GeV/$c$ longitudinally polarized muon beam and NH$_3$
transversely polarized three-cell target. The neighboring target
cells were polarized oppositely which allowed simultaneous
collection of data with both spin polarizations. In order to
minimize the acceptance effects after each 4-5 days polarization was
reversed in all three cells. The kinematic cuts applied in the
analysis are the following ones: $Q^2>1$ (GeV/c)$^2$, $W>5$ GeV,
$0.1<y<0.9$, $P_{hT}>0.1$ GeV/c and $z>0.2$.

The estimator used for the evaluation of the raw asymmetries is
based on an extended unbinned maximum likelihood method and as well
as the whole analysis procedure is exactly the same as the one
applied for already published Collins and Sivers asymmetries on
proton. A more detailed description of the COMPASS spectrometer and
analysis methods can be found in:
\cite{Abbon:2007pq,Alexakhin:2005iw,Ageev:2006da,Alekseev:2008dn,Alekseev:2010rw,Parsamyan:2007ju}

In figures~\ref{fig:1} and \ref{fig:2} we present six TSAs extracted
from COMPASS 2007 data collected on a proton target. The asymmetries
for positive and negative hadrons are plotted as a function of $x$,
$z$, $P_{hT}$ and $W$. The systematic uncertainties have been
estimated separately for each asymmetry for positive and negative
hadrons and are given by the bands. Next, figures~\ref{fig:3} and
\ref{fig:4} show the $x$-dependence of the calculated for COMPASS
kinematical region asymmetries from \cite{Kotzinian:2006dw},
\cite{Pasquini:2008ax} and \cite{Kotzinian:2008fe} together with the
COMPASS measured values. In figure~\ref{fig:3} we show only SSAs for
positive and negative hadrons, while figure~\ref{fig:4} is dedicated
to DSAs. One can see that at maximum UT asymmetries are predicted to
be of order of $1-2\%$ and LTs around $5-10\%$ and that the
theoretical curves basically stay within the error bands showing an
agreement with COMPASS measurements at this level of the statistical
accuracy. Yet, improved precision is needed for definitive
conclusions and further analysis.
\section{Conclusions}\label{concl}

We have presented preliminary results for six new target transverse
spin dependent asymmetries extracted from COMPASS 2007 data
collected on a proton target. These measurement together with the
published Collins and Sivers TSAs on proton and with our deuteron
measurements complete the whole set of eight azimuthal asymmetries
allowed in SIDIS and give another access to the studies of TMD DFs
and FFs. Obviously, presented comparison plots for different theory
predictions show that the statistical accuracy has yet to be
enhanced to draw some decisive conclusions. However, in 2010 COMPASS
collected a new large sample of data with transversely polarized
proton target, and we look forward to significantly improve the
precision of our results.
\section*{References}\label{refs}
\bibliography{iopart-num}
\end{document}